\documentclass[usenatbib]{mn2e}

\usepackage{graphicx}
\usepackage{amssymb}

\newcommand{\aj}{AJ}
\newcommand{\apj}{ApJ}
\newcommand{\apjl}{ApJ}
\newcommand{\apjs}{ApS}
\newcommand{\aap}{A\&A}
\newcommand{\mnras}{MNRAS}

\newcommand{\atel}{ATel}

\newcommand{\apss}{Ap\&SS}

\title[Black hole candidate XTE\,J1752-223: \emph{Swift} observations of canonical states during outburst]
{Black hole candidate XTE\,J1752-223: \emph{Swift} observations of canonical states during outburst}

\author[P.A.~Curran et al.]
{P.A.~Curran$^1$\thanks{Present address: AIM, CEA/DSM - CNRS - Universit\'e Paris Diderot, Irfu/SAP, Centre de Saclay, Baˆt. 709, FR-91191 Gif-sur-Yvette Cedex, France; e-mail: peter.curran@cea.fr},  
T. J. Maccarone$^{2}$,
P.~Casella$^{2}$, 
P.A.~Evans$^{3}$,
W. Landsman$^{4}$, \newauthor
H.A.~Krimm$^{5,6}$,  
C.~Brocksopp$^{1}$,
M.~Still$^{7}$
\\
$^1$Mullard Space Science Laboratory, University College of London, Holmbury St Mary, Dorking, Surrey RH5\,6NT, UK\\
$^2$School of Physics and Astronomy, University of Southampton, Southampton, Hampshire, SO17\,1BJ, UK \\
$^3$Department of Physics and Astronomy, University of Leicester, University Road, Leicester LE1\,7RH, UK \\
$^4$Adnet Systems, NASA/Goddard Space Flight Center, Code 667, Greenbelt MD 20771, USA \\
$^5$NASA/Goddard Space Flight Center, Astrophysics Science Division, Code 661, Greenbelt, MD 20771, USA  \\
$^6$Universities Space Research Association, Columbia, MD 21044, USA \\
$^7$NASA Ames Research Center, Moffett Field, CA 94035, USA \\
 }

\begin{document}

\date{Accepted/ Received;}

\pagerange{\pageref{firstpage}--\pageref{lastpage}} \pubyear{}

\maketitle

\label{firstpage}


\begin{abstract}
We present \emph{Swift} broadband observations of the recently discovered black hole candidate, X-ray transient, XTE\,J1752-223, obtained over the period of outburst from October 2009 to June 2010. From \emph{Swift}-UVOT data we confirm the presence of an optical counterpart which displays variability correlated, in the soft state, to the X-ray emission observed by \emph{Swift}-XRT. The optical counterpart also displays hysteretical behaviour between the states not normally observed in the optical bands, suggesting a possible contribution from a synchrotron emitting jet to the optical emission in the rising hard state. 
We offer a  purely phenomenological treatment of the spectra as an indication of the canonical spectral state of the source during different periods of the outburst. We find that the high energy hardness-intensity diagrams over two separate bands follows the canonical behavior, confirming the spectral states. Our XRT timing analysis shows that in the hard state there is significant variability below 10\,Hz  which is more pronounced at low energies, while during the soft state the level  of variability is consistent with being minimal. 
These properties of XTE\,J1752-223 support its candidacy as a black hole in the Galactic centre region.
\end{abstract}



\section{Introduction}\label{section:introduction}

Low mass X-ray binaries are for the majority of the time in a state of quiescensce with faint or non-detected X-ray emission, though optical or near-infrared (nIR) counterparts may be visible due to emission from the donor star, or possibly the jet, hot spot, or outer accretion disk. They are often only discovered when they enter an active state of outburst when -- powered by an increased level of accretion onto the central, compact object (black hole or neutron star) -- there is a dramatic increase of the X-ray, optical/nIR and radio flux. 
During these outbursts the systems have been observed to go through a number of high energy spectral states before returning to a quiescent state, usually on times scales of weeks, months or even longer. 
These states are a generally low intensity, power-law dominated, {\it hard } state followed by a usually, higher intensity, {\it thermal-dominant}, {\it soft} state which decreases in flux, via a late hard state, over time. Additionally, the hard states are  associated with aperiodic variability of the light curve not present in the soft state  (see \citealt{mclintock2006:csxs157} for a fuller description of the various possible states).

XTE\,J1752-223, a new X-ray transient and  black hole candidate \citep{munoz2010:MNRAS404,markwardt2009ATel.2261,shaposhnikov2009ATel.2269}  in the Galactic center region, was detected on 2009-10-23 at 19:55 UT (MJD 55128.33) by RXTE and on  2009-10-24 at 14:18:50 UT by the BAT instrument on board the \emph{Swift} satellite \citep{markwardt2009ATel.2258}. The high energy, variable emission of the source was confirmed in the following days by \emph{Swift}/XRT \citep{markwardt2009ATel.2261} and RXTE \citep{remillard2009ATel.2265,shaposhnikov2009ATel.2269} as well as by MAXI/GSC  \citep{nakahira2009ATel.2259} and \emph{Fermi}/GBM \citep{wilson2009ATel.2280}. An optical and nIR counterpart was proposed by  \cite{torres2009ATel.2263,torres2009ATel.2268} while a radio source coincident with the X-ray position \citep{markwardt2009ATel.2261} was observed by the Australia Telescope Compact Array (ATCA; \citealt{brocksopp2009ATel.2278}).  The source was observed to have undergone a state transition in mid-January 2010 (MJD $\sim 55210$; \citealt{homan2010:ATel.2387}), from a spectrally hard to a spectrally soft X-ray state. This was identified by a de-coupling of high energy ($\gtrsim 4$\,keV) and low energy ($\lesssim 4$\,keV) light curves \citep{nakahira2010:arXiv1007} 
which had previously traced each other, and by the dominance of a thermal component. The source was observed to have reverted to a hard state at the end of March 2010 (MJD $\sim 55280$; \citealt{Darias2010:ATel.2518}), after the low energy light curve decreased to trace once more the high energy light curve; the thermal component was no longer dominant. It has also been suggested that the  turn-on of a compact jet has been observed in this system \citep{Buxton2010:ATel.2549} during this transition back to the low/hard state.

In this article we present the \emph{Swift} -- Burst Alert Telescope, X-ray Telescope and Ultraviolet/Optical Telescope -- monitoring observations of XTE\,J1752-223, obtained over the period of the outburst. Based on these  data, we identify the periods of the various states and compare the behavior of the major photometric, spectral and timing  parameters during these states to those expected from black hole X-ray binaries.
In section \ref{section:observations} we introduce the observations and reduction methods, while in section \ref{section:analysis} we present the results of our photometric, spectral and timing analyses of the data. We summarise our findings in section \ref{section:conclusion}.


\section{Observations \& Reduction}\label{section:observations}

\emph{Swift} \citep{gehrels2004:ApJ611} observed XTE\,J1752-223 with its narrow field instruments, the Ultraviolet/Optical Telescope (UVOT; \citealt{roming2005:SSRv120}) and the X-ray Telescope (XRT; \citealt{burrows2005:SSRv120}), on 55 epochs from October 25 to November 03 2009 and from February 04 2010 to  June 5 2010. These observations (under Target IDs 31532, 31640 and 31688)  coincide with the period of the outburst. Observations with the narrow field instruments were interrupted for 3 months due to the source becoming constrained by the position of the Sun. Wide field instruments also became sun constrained but at a later date and for a shorter period of time, depending on the satellite and instrument in question (e.g. MAXI was constrained from December 15 to 24). During the period of the outburst, as it had been on previous occasions, the field was also being observed (though constrained from December 13 to 28) by \emph{Swift}'s Burst Alert Telescope (BAT; \citealt{barthelmy2005:SSRv120}) which, in combination with  XRT and UVOT,  affords quasi-simultaneous observations from 0.002 to 150 keV.  
Reduction of \emph{Swift} data was carried out using version 3.4 of the \emph{Swift} software, released as part of HEASOFT 6.7.
All uncertainties are quoted at the $1\sigma$ confidence level.

\subsection{Ultraviolet/Optical Telescope}

UVOT data have been pre-processed at the \emph{Swift} Data Center (see  \citealt{breeveld2010MNRAS}) and require only minimum user processing. As the pointing of \emph{Swift} is only accurate to a few arcseconds, the standard pre-process pipeline provides an astrometric correction by matching sources with the USNO-B catalog.  However, in this case the automatic matching failed because the field differs significantly between the ultraviolet and the USNO-B bandpass. To overcome this, we created our own astrometric reference catalog by selecting blue and bright field stars in the GSC-2.3 catalog \citep{lasker2008:AJ136} to determine astrometric corrections for the data. 

\begin{figure} 
  \centering 
  \resizebox{\hsize}{!}{\includegraphics[angle=0]{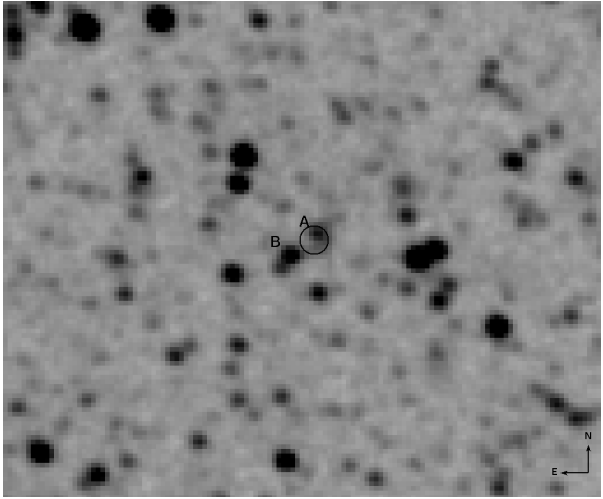} }
  \caption{UVOT 1853\,s $v$ band $130\arcsec \times 100\arcsec$ (MJD 55130-55138) image with XRT $1\sigma$ error circle marked along with optical sources A and B.}
  \label{fig:uvot_v} 
\end{figure}

The image data of each filter, from each observation sequence, i.e., with a given observation ID, were summed using {\tt uvotimsum}. All sequences were then combined in an individual fits file, using  {\tt fappend}, which was again summed using {\tt uvotimsum} to obtain a deep image of the field in each filter. From these images, two possible optical counterparts of XTE\,J1752-223 are identified, consistent with the X-ray position \citep{markwardt2009ATel.2261}; source A, consistent with the counterpart proposed by  \cite{torres2009ATel.2263}, is within $1\sigma$ and source B is within $2\sigma$ of the X-ray position. The positions (J2000; 0.31\arcsec error) of the possible counterparts were derived from such a deep (1853\,s) $v$ band image summed between MJD 55130 and 55138 (Figure \ref{fig:uvot_v}), using the {\tt uvotdetect} command, as: \\
\begin{tabular}{l l l l} 
 & A & 17:52:15.08 &  $-$22:20:32.9  \\
 & B & 17:52:15.52  &  $-$22:20:37.4   \\
\end{tabular}\\
We caution however, that given the crowded nature of this field (see finding chart of \citealt{torres2009ATel.2263}), it is possible that these two sources are in fact unresolved groupings. Though this does not effect our study of source variability, it has a significant impact on the spectral fitting.

Photometry of the two sources in individual sequences are derived via {\tt uvotmaghist}, using an extraction region of radius 4\arcsec and  {\tt XSPEC} compatible spectral files for source A created with the same region, using {\tt uvot2pha}.
Magnitudes are  based on the UVOT photometric system \citep{poole2008:MNRAS383}.   
Source B shows no  variation over the time of the observations, while source A shows significant variability in the $v$, $b$ and $u$ bands (Figure\,\ref{fig:uvot_lc}) and is undetected in the individual UV band ($uw1$, $um2$, $uw2$) images, except for a few $uw1$ detections in late, deep images.  
The average magnitudes of each of these sources, from the deep, summed images between  MJD 55130 and 55138, where the magnitude of source A is relatively constant,   are given in Table\,\ref{table:uvot_mags}.  These are uncorrected for the Galactic extinction in the direction of XTE\,J1752-223  \citep{schlegel1998:ApJ500} of $E_{(B-V)} = 1.554$.  Taking this value as an upper limit to the extinction of source A and using the effective wavelengths of the \emph{Swift} filters \citep{poole2008:MNRAS383} and the parameterization of \cite{pei1992:ApJ395}, the extinctions is the \emph{Swift} bands are: $A_{v} \leq 4.7$,  $A_{b} \leq 6.1$,  $A_{u} \leq 7.6$,  $A_{uvw1} \leq 10.5$, $A_{uvw2} \leq 12.5$, $A_{uvm2} \leq 15.0$, though these should be treated with caution as estimates of the extinction so close to the Galactic plane ($<5\deg$) are unreliable.

\begin{table}	
  \centering	
  \caption{UVOT magnitudes and 3$\sigma$ limits of sources A and B as well as the background limit of source A, $limit$, from summed images taken  between  MJD 55130 and 55138. Uncorrected for interstellar extinction of $E_{(B-V)} \lesssim 1.554$.}
\label{table:uvot_mags} 	
\begin{tabular}{l l l l l } 
  \hline          
  Band & Exp (s) &  A      & B   &  $limit$ \\
  \hline
  $v$  & 1853  &  16.79 $\pm$ 0.03  &  16.20 $\pm$ 0.03    & 19.9  \\   
  $b$  & 703   &  18.10 $\pm$ 0.07  &  17.57 $\pm$ 0.05    & 20.3  \\   
  $u$  & 1042  &  18.37 $\pm$ 0.05  &  18.29 $\pm$ 0.08    & 20.3  \\   
  $uw1$ & 1098  &  19.53 $\pm$ 0.18  &  19.28 $\pm$ 0.19    & 20.5  \\ 
  $um2$ & 1504  &  $>$20.8  &  $>$20.7            & 20.8  \\   
  $uw2$ & 2083  &  20.9 $\pm$ 0.3  &  $>$21.0      & 21.1  \\   
  \hline 
\end{tabular}
\end{table}

\subsection{X-ray Telescope}

\begin{figure} 
  \centering 
  \resizebox{\hsize}{!}{\includegraphics[angle=0]{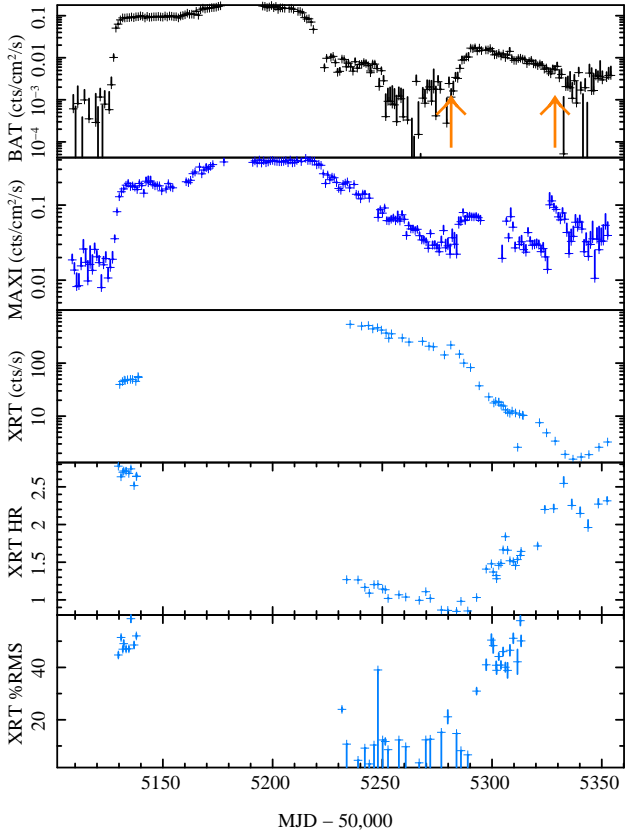} }
  \caption{High energy (\emph{Swift}-BAT [15-150\,keV], MAXI [4-10\,keV], \emph{Swift}-XRT [0.3-10\,keV]) light curves during the period of outburst. 
Also shown are the XRT  hardness ratio, HR  [1.5-10\,keV / 0.3-1.5\,keV] and percentage RMS. The absence of data around MJD $\sim$55180 is due to the various instruments becoming sun-constrained. The  arrows signify the dates between which the source is transitioning from the soft state to the late, hard state (MJD 55280, 55328).}
  \label{fig:xray_lc} 
\end{figure}

The XRT Windowed Timing mode  (WT; \citealt{hill2004:SPIE5165}) event data [0.3-10\,keV] were analysed using the tools presented by \cite{evans2009:MNRAS397}. These correct for pile up and the affect of bad CCD columns. Version v011 response matrices were used for the spectral analysis.
The best XRT position of 17:52:15.14 $-$22:20:33.8  (J2000; 3.1\arcsec error)  is as previously reported \citep{markwardt2009ATel.2261}. 
Background subtracted XRT count rate light curves are binned per observing sequence (Figure\,\ref{fig:xray_lc}; combined with MAXI  observations). Unbinned source light curves (not background subtracted), for each sequence, were used to produce power density spectra (PDS) and to calculate the RMS variability with the FTOOLs, {\tt powspec} and {\tt lcstats} respectively. Within each of these the binning was set to be N times the minimum bin size (1.766\,ms) where N=1 for the PDS and N=8 (14.128\,ms) for the RMS variability, giving frequency ranges of 0.02-570\,Hz and 0.02-70\,Hz respectively.
The average PDS were calculated, within  {\tt powspec}, from an input list of the light curves of the required sequences, using the same binning.
Lags between the standard XRT low [0.3-1.5\,keV] and high [1.5-10\,keV] energy band unbinned light curves were calculated with {\tt crosscor} and a bin size set to N=8 (14.128\,ms).

\subsection{Burst Alert Telescope}

BAT detector plane histograms (DPHs)  were obtained in survey mode; these are 2 dimensional images with full spectral data and coarse time  binning.
Spectra and light curves were extracted from these DPHs using mask weighting  (also known as ray-tracing) to account for the position of the source on the coded aperture  mask.
The data were initially processed with the FTOOL, {\tt batsurvey} which applies data quality cuts and uses the tools {\tt batfftimage} and {\tt batcelldetect} to extract 8-channel spectra for the source.  We then used custom software to format the spectra for {\tt XSPEC}, and {\tt batdrmgen} to generate the response files.  The light curve was derived using a similar method.
BAT, as a survey telescope, had been observing the field of XTE\,J1752-223 since the start of \emph{Swift} operations in January 2006, though not monitoring the source. The extracted BAT light curve (15-150\,keV) of the source from January 12 2006 through to the present, displays no activity until the 2009 outburst.

\begin{figure} 
  \centering 
  \resizebox{\hsize}{!}{\includegraphics[angle=-90]{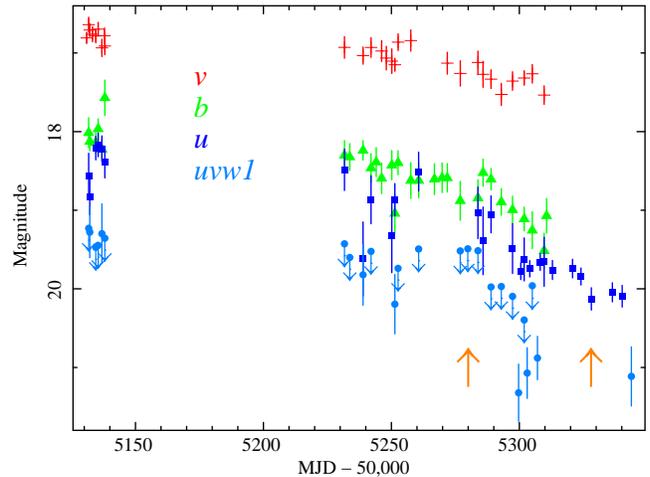} }
  \caption{UVOT  $v$ (crosses), $b$ (triangles), $u$ (squares) and $uw1$ (circles with $3\sigma$ limits) band light curves for source A, which show variability; dimming by 1 magnitude or more from the start to the end of observations. The up-pointing arrows signify the dates between which the source is transitioning from the soft state to the late, hard state (MJD 55280, 55328).}
  \label{fig:uvot_lc} 
\end{figure}

\section{Analysis \& Discussion}\label{section:analysis}

\subsection{Energy Spectra}\label{section:SEDS}

The joint XRT-BAT spectra were fit within {\tt XSPEC (12.5.1)}, using C-statistics and interstellar absorption modeled by {\tt tbabs}; initially we fit an absorbed power law ({\tt pow}) to the spectra before adding a thermal component ({\tt diskbb}). We note that these are phenomenological models for the respective spectral states, not a suggestion that these are necessarily the best physical models; they should not be used to deduce physical constraints but only as an indication of the spectral state of the source. 

\begin{figure} 
  \centering 
  \resizebox{\hsize}{!}{\includegraphics[angle=-90]{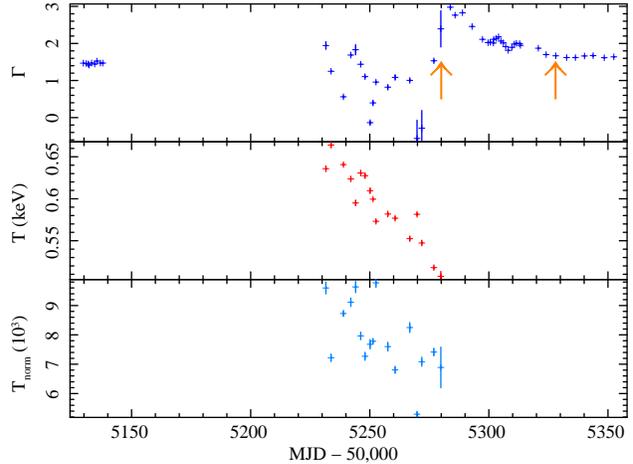} }
  \caption{Best fit parameters of the joint XRT-BAT spectral fits: power law photon index, $\Gamma$,  thermal component temperature, $T$, and thermal component normalization, $T_{{\rm norm}}$, in units of $10^{3}$. 
Note that the photon indices, in the period where there is a thermal component, are not reliable. 
The absence of data around MJD $\sim$55180 is due to the various instruments becoming sun-constrained. The arrows signify the dates between which the source is transitioning from the soft state to the late, hard state (MJD 55280, 55328).}
  \label{fig:spectral} 
\end{figure}

The early data (MJD 55131-55138) are well described by the power law of average photon index,  $\Gamma = 1.46 \pm 0.03$, (Figure\,\ref{fig:spectral}) and do not require a thermal component - clearly indicating that the source is in the hard state. The absorption of a simultaneous fit of these sequences is $N_{{\rm H}} = 0.513 \pm 0.003 \times 10^{22}$\,cm$^{-2}$, greater than the Galactic value of  $N_{{\rm H}}  = 0.46 \times 10^{22}$\,cm$^{-2}$ \citep{kalberla2005:A&A440} but lower than the $N_{{\rm H}} = 0.72^{+0.01}_{-0.04} \times 10^{22}$\,cm$^{-2}$ found by \citet{munoz2010:MNRAS404}. Data observed from MJD 55234-55280 additionally support a soft state thermal component decreasing from a temperature of $\sim 0.7$\,keV to $\sim 0.5$\,keV over the period, as well as decreasing normalisation. Due to the degeneracy between the power law and the thermal components and lack of counts from the BAT, the power law index is poorly constrained in a manner not obvious from the nominal fit errors but causing a high level of variability in the upper panel of Figure\,\ref{fig:spectral}; these values are not reliable.

From MJD 55283 the data no longer support a thermal component, though the signal to noise at these low counts rates cannot rule out a thermal component. Additionally, as can be seen from Figure\,\ref{fig:spectral}, the power law indices are decreasing until MJD 55328, signaling the transitional period between the soft and hard states, in agreement with the increasing hardness ratio (Figure\,\ref{fig:xray_lc}). 
From MJD 55328 to the end of our observations, on MJD 55352, the spectra are  described by a power law of relatively constant photon index,  the average of which is $\Gamma = 1.64 \pm 0.02$. The power law indices and the hardness ratios at these late times fail to return to the original, rising hard state values ($\Gamma = 1.46 \pm 0.03$), indicating that, though the count rate is lower than it was in that rising hard state,  the source has not yet returned to its original hard state or that the late hard state has marginally different spectral properties.

\subsection{Optical Counterpart}\label{section:optical}

\begin{figure} 
  \centering 
  \resizebox{\hsize}{!}{\includegraphics[angle=-90]{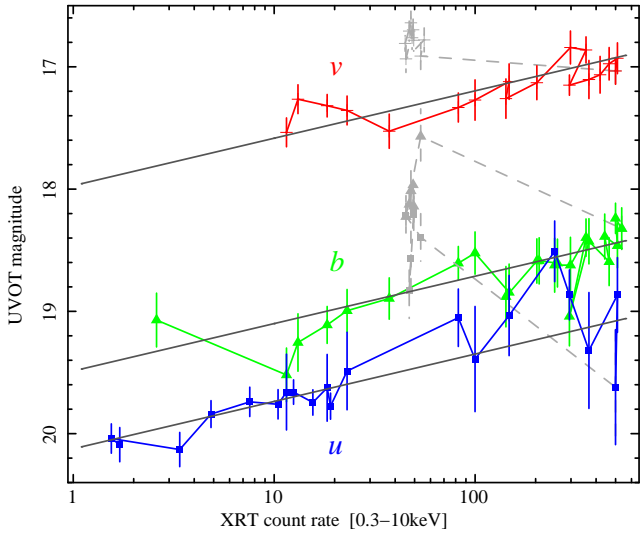} }
  \caption{ UVOT magnitude in three filters versus XRT count rate. Data points in grey indicate observations before MJD $\sim 55230$, i.e., the rising hard state. The solid black lines represent the simultaneous power law fit to the data after MJD $55230$.}
  \label{fig:uvot-xrt} 
\end{figure}

An optical and nIR counterpart was proposed by  \cite{torres2009ATel.2263,torres2009ATel.2268} based on a comparison between their images and those from the Digitized Sky Survey, 2MASS and UKIDSS. We have confirmed UVOT source A as this optical counterpart on the basis of a significant, though low amplitude  variability in the $v$, $b$ and $u$ bands (Figure\,\ref{fig:uvot_lc}). Furthermore, these optical magnitudes, or count rates, exhibit a power-law correlation with the X-ray count rate, as observed by XRT at times after MJD 55232 (Figure\,\ref{fig:uvot-xrt}), i.e., when the source is in the soft state and later, during the transition to the late hard state. 
The earlier, hard state data does not fit on this correlation, but displays hysteretical behaviour and for the given X-ray count rate it spans has a significantly higher magnitude in all bands. 
This is similar to the hysteretical behaviour observed in the nIR for a number of transients \citep{russell2007:MNRAS379}, most notably XTE\,J1550-564,   where the additional hard state emission is attributed to optically thin synchrotron emission from a jet \citep{jain2001:ApJ554,homan_2005ApJ...624,hynes_2006ApJ...651}, which would be quenched in the soft state (e.g., \citealt{fender2004:MNRAS355}), and weak at low X-ray luminosities, leading to the hysteresis. 
Though this hysteresis is not normally observed in the optical bands, the optical data presented here is in good agreement with the synchrotron emitting jet making a significant contribution to the optical emission in the rising hard state. 
While the observed $v$ band brightening from MJD 55292 corresponds to the suggested turn-on, in nIR, of a compact jet between MJD 55293 and 55298 \citep{Buxton2010:ATel.2549}, it is not corroborated by the other bands and is not statistically significant, though it may indicate the turn-on of a jet and the transition to the hard state; at such low X-ray luminosity levels, the jet contribution to the optical would be minimal.

From a simultaneous fit of the  $v$, $b$ and $u$ bands the correlation is found to be ${\rm Rate}_{{\rm optical}} \propto {\rm Rate}_{{\rm X-ray}}\,^{0.16\pm0.01}$ with observed  colours of $u-v = 2.15 \pm 0.09$ and $u-v = 1.43 \pm 0.11$  ($\chi^2_{\nu}=1.05$), though the individual bands display indices of $0.10\pm0.04$, $0.18\pm0.03$ and $0.20\pm0.02$ respectively, hinting at possible spectral evolution. 
Such a correlation is more usually associated with the hard state \citep{russell2006:MNRAS371} but we cannot say if the early, hard state data follows another correlation (expected to be ${\rm Rate}_{{\rm optical}} \propto {\rm Rate}_{{\rm X-ray}}\,^{0.6}$; \citealt{russell2006:MNRAS371}) due to the very limited range of magnitudes and X-ray count rates over this period. 
Though we find a correlation between the optical and X-ray, we cannot imply causality or say which leads without a study of the phase lag between the two bands, which is not possible given the time resolution of the optical data. Similarly a timing analysis to investigate variability of the optical data is not possible.  The correlation may be due to reprocessing of the X-rays or through both originating from the accretion disk; it is also possible that optical emission in the soft state originates from a jet, the state of which traces the spectral states \citep{fender2004:MNRAS355. }

The quality of the UVOT spectral data does not support any spectral evolution in the optical, nor does it enable us to specify the most appropriate spectral model for the source such as a power law associated with the above suggested jet emission. Hence, we model the data phenomenologically in {\tt XSPEC}, by a power law (${\rm Flux} \propto \nu^{\beta}$) and a dust extinction component ({\tt zdust}), though we note that this is not a particularly good fit. Given that an unknown amount of Galactic extinction, $E_{(B-V)} \leq 1.554$  \citep{schlegel1998:ApJ500}, effects the data we can only place limits on the spectral index of $ -5.3 \lesssim \beta_{{\rm optical}} \lesssim 2.0$, though the nominal best fit is at $\beta_{{\rm optical}} = -1.6 \pm 0.2$ for $E_{(B-V)} = 0.76 \pm 0.04$.  Given the uncertainties of the spectral slope and the lack of phase lag information, we are unable to make a meaningful comparison to the predictions of the various models.

\subsection{Hardness-Intensity Diagrams}\label{section:HIDs}

\begin{figure}
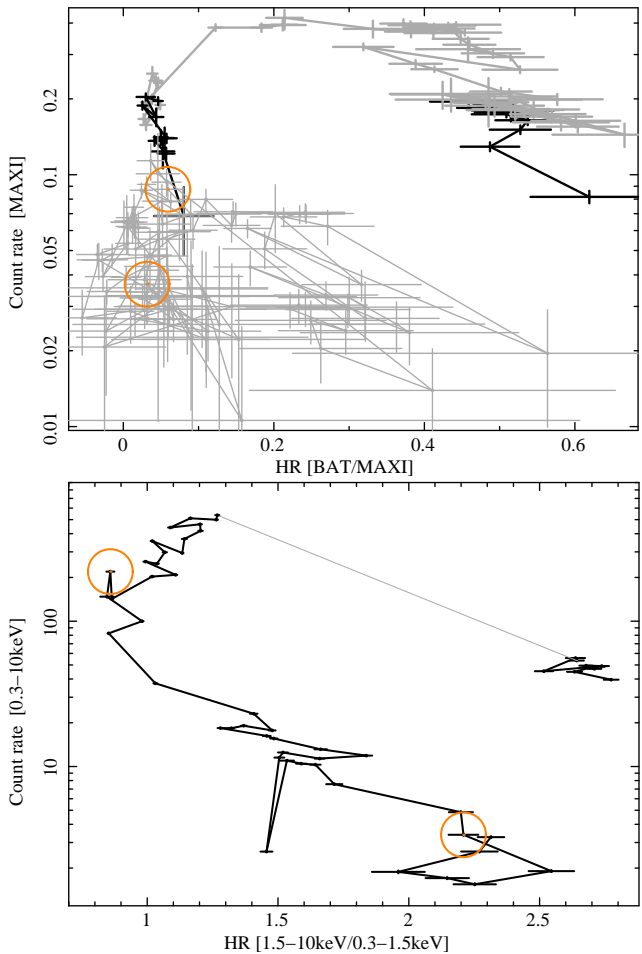
 
  \centering 
  \resizebox{\hsize}{!}{\includegraphics[angle=-90]{fig.HID-BAT_MAXI.ps} }  
  \resizebox{\hsize}{!}{\includegraphics[angle=-90]{fig.HID-XRT.ps} } 
\caption{Hardness-Intensity diagrams (HID) for both BAT/MAXI ({\it upper}) and XRT ({\it lower}). The grey lines, at high count rates in the XRT and BAT plots show the time during which the XRT was unable to observe (MJD 55139-55233),  while the circles show the start and end of the transition from the soft state at MJD 55280 and 55328.  Note that at low count rates ($\lesssim 0.1$) the BAT/MAXI HR is not reliable; the later, low count rate, data points are shaded for clarity of the plot.}
 \label{fig:HID} 
\end{figure}

The Hardness-Intensity diagrams (HID) for both the XRT data [1.5-10\,keV / 0.3-1.5\,keV] and the quasi-simultaneous (same day)  BAT/MAXI data starting at the same time are shown in Figure\,\ref{fig:HID}. Due to the fact that the BAT/MAXI data are only quasi-simultaneous, the hardness ratio (HR) of BAT [15-150\,keV] with respect to MAXI [4-10\,keV] at low BAT count rates ($\lesssim 0.01$), where the relative variability over a day is greater, are underestimated by the error bars. 
In general,  HIDs follow a canonical trajectory (e.g., \citealt{homan2001:ApJS132,homan_2005ApSS...300,belloni_2010LNP...794}):
increasing intensity at a high HR to a maximum in the hard state, indicating the power law shape of the spectra; decrease of HR at a high intensity through an intermediate state as the thermal component of the spectra becomes more dominant; decrease of intensity at the low HR level during the soft state; before the HR increases at low intensity, through an intermediate state, to the original value.
This is consistent with both the XRT data, though a significant amount of the trajectory in the first intermediate state is missing  due to the source being sun constrained for a period of 3 months, and the BAT/MAXI data in which the final phase is obscured due to the large error bars on the HR. It is clear from the XRT HID that the hardness has yet to increase to the original value, indicating that the source has not yet returned to its original hard state, though the count rate is lower than it was in that rising hard state, or that the late hard state has different properties. These tracks are common among black hole binary sources --  though are also observed in neutron star binary sources such as Aquila\,X-1 (e.g., \citealt{tudose_2009MNRAS.400,millerjones_2010ApJ...716})  -- and further confirm the spectral state of the system over the periods in question.

\subsection{Timing Analysis}\label{section:timing}

The RMS variation of the light curves exhibit a similar evolution to hardness ratio (Figure\,\ref{fig:xray_lc}), from average RMS $=53.9 \pm 0.3$ per cent in the observed period of the rising hard state (MJD 55131-55138) to $\lesssim 12$ percent during the soft state observations and increasing to the original levels at later times. Examining the light curves in the standard XRT low [0.3-1.5\,keV] and high [1.5-10\,keV] energy bands we find that, during the rising hard state this variability is more pronounced in the low band with an average ratio between the two of $0.84 \pm 0.02$. This suggests an energy dependency of RMS variation, in contrast to  \citet{munoz2010:MNRAS404} who find a flat RMS spectrum \citep{vaughan2003MNRAS.345,Gierlinski_2005:MNRAS363} from XTE data at energies in the range  2-10\,keV. The lag between the two bands, calculated via a cross correlation, is consistent with zero in all cases though is not tightly constrained.

\begin{figure} 
  \centering 
  \resizebox{\hsize}{!}{\includegraphics[angle=0]{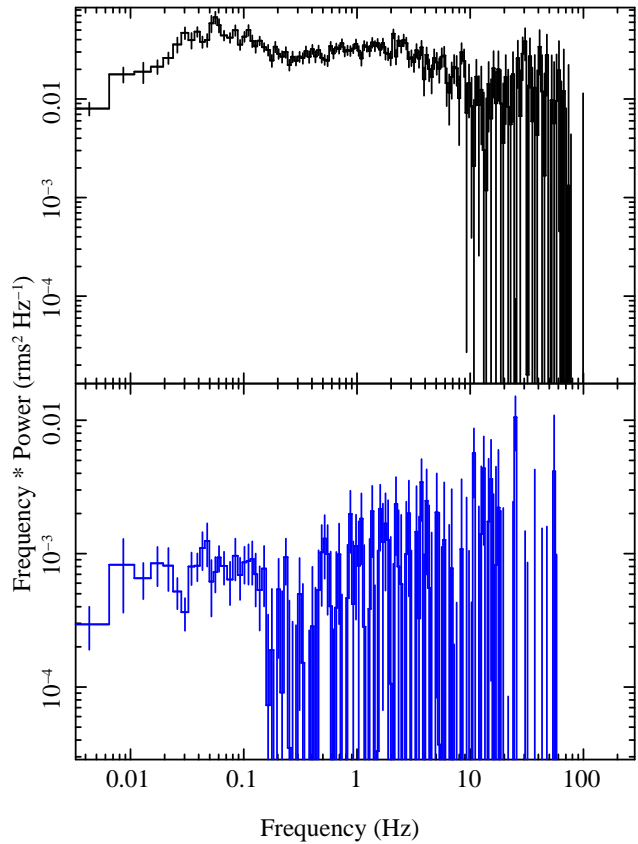} }
  \caption{The average power density spectrum (PDS) for XRT light curve in the hard  state (MJD 55130-55138; {\it upper}) exhibits the aperiodic variability of the light curve not present in the average soft state PDS (MJD 55234-55289; {\it lower}).}
  \label{fig:powspec} 
\end{figure}

\begin{figure} 
  \centering 
  \resizebox{\hsize}{!}{\includegraphics[angle=-90]{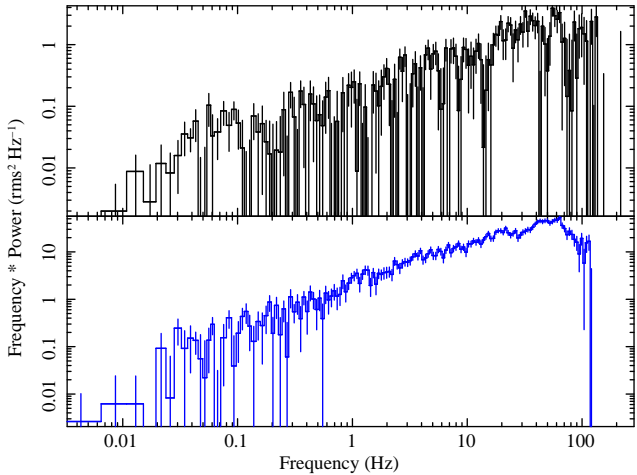} }
  \caption{The power density spectra (PDS) for the XRT source light curve,  at low count rates ({\it upper}; example observations on MJD 55340 where the count rate is 1.7 cts/s), is heavily affected by that of the  background light curve at the same time ({\it lower}).}
  \label{fig:powspec_back} 
\end{figure}

Examining the average power density spectra of the states (Figure\,\ref{fig:powspec}) we see that the hard state has a high level of power in low frequency ($\lesssim 10$\,Hz) variability, though there is no significant sign of quasi-periodic oscillations (QPOs).  The soft state, on the other hand, has a much lower lever of power and only at frequencies $\lesssim 0.2$\,Hz, as is manifested by the $\lesssim 12$ percent limit of RMSs during this period  (MJD 55234-55289).
After MJD 55321, where the XRT count rate drops below 10 cts/s, the light curves and PDSs are heavily affected by background counts (Figure\,\ref{fig:powspec_back}) and hence the RMS values for these observations are unreliable and not plotted in Figure\,\ref{fig:xray_lc}.

The excess of low energy variability in the rising hard state is similar to the behaviour of the hard state variability in SWIFT\,J1753.5-0127 and GX\,339-4  as reported by \cite{wilkinson_2009MNRAS.397}.  
Within the framework of their model, enhanced low energy variability (at frequencies $\lesssim 0.1$\,Hz) is due to a contribution from the thermal emission of the accretion disk. The excess variability is attributed to intrinsic  disk variability --  due to accretion rate fluctuations -- as opposed to reprocessing of the spectral power-law component of the emission.  
Though a thermal component is not supported by the spectra of XTE\,J1752-223 in the rising hard state, the excess variability we observe suggests that one might contribute to the flux at low energies. More detailed spectral fits using the methods introduced in \cite{wilkinson_2009MNRAS.397} are required to test this possibility. Data, over a greater energy range, which better constrains column density and photon index may also be necessary.

\section{Conclusion}\label{section:conclusion}

XTE\,J1752-223 was observed by \emph{Swift} in a relatively low intensity, hard state from 2009-10-26 to 2009-11-03 (MJD 55131-55138) which was signified by hard power law energy spectra, and high levels ($\sim 50$ per cent) of RMS variability in the X-ray light curve. XTE\,J1752-223 then became  constrained by the position of the Sun and was unobservable until 2010-02-04  (MJD 55234) by which time it had experienced a ten fold increase in the flux observed by XRT [0.3-10.0\,keV], which proceeded to decrease to the original levels over a period of 2 months. The spectra during this period clearly exhibited a strong thermal component and the variability and hardness ratios of the light curves had dropped significantly, consistent with the source being in a soft state until 2010-03-25 (MJD 55280). 
Observations from 2010-03-28 (MJD 55283) to 2010-05-12 (MJD 55328) do not support a thermal component but do display a decreasing photon index along with increasing hardness and RMS variability. This suggests a transitional state before reverting to the hard state, with a relatively constant photon index, in which the system is observed until the end of observations on 2010-06-05 (MJD 55352). 
The high energy hardness-intensity diagrams over two separate bands follow the canonical behavior associated with a black hole binary, further confirming the spectral states of the system during the outburst.
Our timing analysis shows that in the hard state there is significant variability below 10\,Hz  which is more pronounced at low energies, while during the soft state the level  of variability is consistent with being minimal.

We are able to confirm that the UVOT source at the position of the proposed optical and near-infrared counterpart is associated with XTE\,J1752-223 due to the observed low amplitude variability and, in the soft state and later, correlation with the X-ray emission as measured by XRT.  
However, we cannot state that the optical flux is causally connected to the X-ray emission via reprocessing of the X-rays or through both originating from the accretion disk; given the uncertainties of the spectral slope and the lack of phase lag information, we are unable to make a meaningful comparison to the predictions of the various models. 
The optical counterpart also displays hysteretical behaviour not normally observed in the optical bands; for a given X-ray count rate, the magnitude in the rising hard state is significantly higher than that in the soft state. 
This is similar to the hysteretical behaviour observed in the nIR for a number of transients  where the additional hard state emission is attributed to optically thin synchrotron emission from a jet, which would be quenched in the soft state and weak at low X-ray luminosities. 
Though this hysteresis is not normally observed in the optical bands, it is in good agreement with the synchrotron emitting jet making a significant contribution to the optical emission in the rising hard state.


\section*{Acknowledgements}
 
We thank the referee for constructive comments and Paul Kuin for useful discussions on UVOT.  
PAC, PAE and CB  acknowledge support from STFC.
TJM thanks the EU FP7 for support through grant number ITN 215212 `Black Hole Universe'.
PC acknowledges funding via a EU Marie Curie Intra-European Fellowship under contract no. 2009-237722. 
This research has made use of: 
Swift data supplied by the UK Swift Science Data Centre at the University of Leicester;  
and MAXI data provided by RIKEN, JAXA and MAXI teams.

\label{lastpage}

\end{document}